\documentclass[aps,pre,twocolumn,footinbib,longbibliography,superscriptaddress]{revtex4-1}
\usepackage{amsmath}
\usepackage{amsfonts}
\usepackage{amssymb}
\usepackage{lmodern,dsfont}
\usepackage{graphicx}
\usepackage[usenames,dvipsnames]{xcolor}
\usepackage{bm}
\usepackage{blkarray}
\usepackage{blindtext}
\usepackage[english=american]{csquotes}

\newcommand{\kb}{k_\text{B}}
\newcommand{\Av}[1]{\left\langle #1 \right\rangle}
\newcommand{\av}[1]{\langle #1 \rangle}
\newcommand{\n}{\nonumber}
\newcommand{\nn}{\nonumber \\}

\renewcommand{\eqref}[1]{Eq.~(\ref{#1})}

\begin{document}

\title{Fluctuation-response inequality out of equilibrium}

\author{Andreas Dechant}
\affiliation{WPI-Advanced Institute for Materials Research (WPI-AIMR), Tohoku University, Sendai 980-8577, Japan}
\author{Shin-ichi Sasa}
\affiliation{Department of Physics \#1, Graduate School of Science, Kyoto University, Kyoto 606-8502, Japan}
\date{\today}

\begin{abstract}
We present a new approach to response around arbitrary out-of-equilibrium states in the form of a fluctuation-response inequality (FRI).
We study the response of an observable to a perturbation of the underlying stochastic dynamics.
We find that magnitude of the response is bounded from above by the fluctuations of the observable in the unperturbed system and the Kullback-Leibler divergence between the probability densities describing the perturbed and unperturbed system.
This establishes a connection between linear response and concepts of information theory.
We show that in many physical situations, the relative entropy may be expressed in terms of physical observables.
As a direct consequence of this FRI, we show that for steady state particle transport, the differential mobility is bounded by the diffusivity.
For a \enquote{virtual} perturbation proportional to the local mean velocity, we recover the thermodynamic uncertainty relation (TUR) for steady state transport processes.
Finally, we use the FRI to derive a generalization of the uncertainty relation to arbitrary dynamics, which involves higher-order cumulants of the observable.
We provide an explicit example, in which the TUR is violated but its generalization is satisfied with equality.
\end{abstract}

\maketitle

Linear response theory is one of the most universal results in physics, from electrodynamics and solid-state physics to quantum mechanics and thermodynamics \cite{Kub57,Han82,Gro85,Bar87,Bar89}. It provides the link between the measured response of a physical system to a small perturbation and the properties of the unperturbed system. For a system in thermal equilibrium, this link takes the form of the fluctuation-dissipation theorem (FDT) \cite{Web56,Kub66}, which states that the response of the system can be characterized through its equilibrium fluctuations. The fluctuations are expressed through equilibrium correlations between the observable and the change in the system's energy induced by the perturbation. The FDT has since been generalized to out-of-equilibrium situations, most notably non-equilibrium steady states \cite{Han82,Har05,Spe06,Pro09,Bai09,Sei10}, where the response to a small time-dependent perturbation is expressed in terms of fluctuations in the steady-state.

Another result of equally universal character is the second law of thermodynamics. During any operation on the system, the total entropy of a system and its environment can never decrease. This is a consequence of the lack of information about the precise microscopic state of the system and environment \cite{Jay65}. This lack of information can be made explicit by introducing randomness and describing the evolution of observables as a stochastic process. In this context, the increase in entropy is due to the irreversibilty of the stochastic dynamics and the change in physical entropy is expressed as the relative entropy between the probability of the system following the time-forward or time-reversed evolution \cite{Sei12}. The relative entropy (or Kullback-Leibler divergence) makes explicit the connection between information and physical entropy; the increase of the latter is a consequence of the mathematical properties of the former.

In this article, we establish a connection between the response of a physical observable and relative entropy in the form of the Kullback-Leibler divergence. 
Our first main result is a fluctuation-response inequality (FRI) for arbitrary stochastic dynamics. 
The statement of the FRI is that the response of the observable is dominated by the Kullback-Leibler divergence between the perturbed and unperturbed probability densities. 
In particular, if the perturbation is weak enough that it allows for a linear response treatment, we obtain our second main result, the linear-response FRI (LFRI). 
This states that the response of any observable is bounded in magnitude from above by the product of the fluctuations of the observable in the unperturbed state and the Kullback-Leibler divergence. 
As our third main result, we show that the LFRI provides a strong constraint on particle transport: the magnitude of the differential mobility is bounded from above by the diffusivity. 
The LFRI thus predicts a universal relation between two physical observables. Another class of such such relations, the recently derived thermodynamic uncertainty relations \cite{Bar15,Gin16,Pie17,Hor17,Dec17}, arises naturally from the LFRI for a specific choice of the perturbation. 
Using the FRI, we generalize these uncertainty relations to arbitrary dynamics, which constitute our fourth main result, and discuss how to understand them from the point of view of symmetries of the observable. %For systems close to equilibrium, we show that any current can be bounded by its equilibrium fluctuations, extending a recent result \cite{Mac18} to time-dependent driving.

\section{Fluctuation-response inequality} 
We consider a general physical system, described by a probability density $P^a(\omega)$, where $\omega$ denotes the degrees of freedom of the system. For example, we may take $\omega = \bm{x}$ to be the coordinates of a diffusing particle and $P^a(\omega)$ the probability density of the particle's position at time $t$. However, we may equally well take $\omega = \lbrace \bm{x}(t)\rbrace_{t \in [0,\mathcal{T}]}$ to be the path traveled by the particle during the time interval $[0,\mathcal{T}]$, in which case $P^a(\omega)$ is a path probability density. We further specify an observable $r(\omega)$. Depending on $\omega$, such an observable may e.~g.~be a function of the particle's position, or the entropy production along the path. We denote the average of $r(\omega)$ by $\av{r}^a = \int d\omega \ r(\omega) P^a(\omega)$. Since $\omega$ is a random variable, the observable $r(\omega)$ will likewise fluctuate. We can characterize the fluctuations of $r(\omega)$ by its deviations from the average $\Delta r(\omega) = r(\omega) - \av{r}^a$. The statistics of those fluctuations are encoded in the cumulant generating function
\begin{align}
K_{\Delta r}^a(h) = \ln \Av{e^{h (r - \av{r}_a)}}^a = K_{r}^a(h) - h \av{r}_a ,
\end{align}
where $K_r^a(h)$ is the cumulant generating function of the observable $h$.
The $n$-th cumulant $\kappa_{n,r}$ of the fluctuations is obtained as the $n$-th derivative of this function with respect to $h$, $\kappa_{n,r}^a = \partial_h^n K_{\Delta r}^a(h) \vert_{h = 0}$. 
The first cumulant is zero, while the second cumulant is the variance $\av{\Delta r^2}^a$. 
We now perturb the system, e.~g.~by applying an external force. 
The perturbation changes the probability density $P^b(\omega)$ and the average of the observable to $\av{r}^b$. 
We refer to $a$ and $b$ as the reference and perturbed system, respectively.
Provided that $P^a(\omega)$ and $P^b(\omega)$ have the same support (i.~e.~$P^a(\omega)/P^b(\omega)$ is finite for all $\omega$), we can write the cumulant generating function as
\begin{align}
K_{\Delta r}^a(h) = \ln \bigg( \int d\omega \ e^{h (r(\omega) - \av{r}^a)} \frac{P^a(\omega)}{P^b(\omega)} P^b(\omega) \bigg).
\end{align}
Since the logarithm is a concave function, we can apply the Jensen inequality to obtain
\begin{align}
K_{\Delta r}^a(h) &\geq \int d\omega \ \ln \bigg(e^{h (r(\omega)-\av{r}^a)} \frac{P^a(\omega)}{P^b(\omega)} \bigg) P^b(\omega) \nn
&= h \big(\av{r}^b - \av{r}^a \big) - D_\text{KL}^{b \Vert a}, \label{cgf-bound}
\end{align}
where $D_{KL}^{b \Vert a}$ denotes the Kullback-Leibler (KL) divergence \cite{Kul51} between the probability distributions
\begin{align}
D_\text{KL}^{b \Vert a} &= \int d\omega \ P^b(\omega) \ln \bigg(\frac{P^b(\omega)}{P^a(\omega)} \bigg) \label{KL-divergence} .
\end{align}
The KL divergence is positive and vanishes only if the probability distributions describing the perturbed and reference system are identical almost everywhere.
It can be regarded as an information-theoretic measure of distance between the two probability distributions, however, it is not a distance in the strict sense, since it is not symmetric and does not satisfy the triangle inequality.
%Taking the average out of the cumulant generating function, we can write
%\begin{align}
%K_r^a(h) = h \av{r}^a + \ln \Av{e^{h (r-\av{r}^a)}}^a \equiv h \av{r}^a + K_{\Delta r}^a(h)
%\end{align}
%and thus the above inequality can be written as
%\begin{align}
%h \big(\av{r}^b - \av{r}^a \big) \leq K_{\Delta r}^a(h) + D_{\text{KL}}^{b \Vert a} .
%\end{align}
Defining $\sigma = \text{sign}(\av{r}^b - \av{r}^a)$, we can write this inequality as
\begin{align}
\big\vert \av{r}^b - \av{r}^a \big\vert \leq \inf_{h > 0} \frac{1}{h} \Big( K_{\Delta r}^a(h \sigma) + D_{\text{KL}}^{b \Vert a} \Big) , \label{relent-bound}
\end{align}
where we can take the infimum since the inequality holds for all values of $h$ for which the cumulant generating function is defined.
In general, the optimal value $h^*$ which yields the tightest bound depends on the explicit functional form of the cumulant generating function.
However, the infimum may be computed explicitly if the distribution of $r$ in the reference system $a$ is Gaussian,
\begin{align}
P_r^a(r) = \frac{1}{\sqrt{2 \pi \av{\Delta r^2}^a}} \exp\bigg[-\frac{(r - \av{r}^a)^2}{2\av{\Delta r^2}^a} \bigg] .
\end{align}
In this case, only the first two cumulants are non-zero and the cumulant generating function of the fluctuations is given by
\begin{align}
K_{\Delta r}^a(h) = \frac{h^2}{2} \av{\Delta r^2}^a .
\end{align}
Minimizing the right-hand side of \eqref{relent-bound} with respect to $h$, we obtain
\begin{align}
h^* = \sqrt{\frac{2 D_{\text{KL}}^{b \Vert a}}{\av{\Delta r^2}^a}} \label{h-gauss}
\end{align}
and thus the inequality
\begin{align}
\big\vert \av{r}^b - \av{r}^a \big\vert \leq \sqrt{2 D_{\text{KL}}^{b \Vert a} \av{\Delta r^2}^a} . \label{gauss-ineq}
\end{align}
This inequality provides a simple relation between the response of the observable $r$ to the perturbation, its fluctuations in the reference system and the information-theoretic distance between the perturbed and reference system.
Even if the distribution of $r$ is not Gaussian, \eqref{h-gauss} still captures the correct scale of the optimal value of $h$.
Writing $h = \alpha h^*$, we obtain
\begin{align}
\big\vert \av{r}^b &- \av{r}^a \big\vert \leq \sqrt{2 D_{\text{KL}}^{b \Vert a} \av{\Delta r^2}^a} \\
& \quad \times \inf_{\alpha > 0} \frac{1}{\alpha} \Bigg[ \frac{1}{2} + \frac{1}{2 D_{\text{KL}}^{b \Vert a}} K_{\Delta r}^a \Bigg( \sqrt{\frac{2 D_{\text{KL}}^{b \Vert a}}{\av{\Delta r^2}^a}} \sigma \alpha \Bigg) \Bigg] \n .
\end{align}
The structure of the right-hand side becomes clearer when writing it explicitly in terms of the cumulants,
\begin{align}
&\big\vert \av{r}^b - \av{r}^a \big\vert \leq \sqrt{2 D_{\text{KL}}^{b \Vert a} \av{\Delta r^2}^a}  \label{non-gauss-ineq} \\
& \ \times \inf_{\alpha > 0} \Bigg[ \frac{1}{2 \alpha} + \frac{\alpha}{2} + \sigma \sum_{n = 3}^\infty \frac{(\sigma \alpha)^{n-1}}{n!} \big(2 D_\text{KL}^{b \Vert a}\big)^{\frac{n-2}{2}} \frac{\kappa_{n,r}^a}{\big(\av{\Delta r^2}^a \big)^{\frac{n}{2}}} \Bigg] \n .
\end{align}
The expression in square brackets is always positive.
In the Gaussian case, the higher-order terms vanish and thus the infimum is realized for $\alpha = 1$, yielding \eqref{gauss-ineq}.
The interpretation of the inequality \eqref{non-gauss-ineq} is the following: The KL-divergence measures the distinguishability of the probability distributions of the reference system $a$ and the perturbed system $b$. 
Thus, it is natural to expect that the change in the expectation of any observable from $a$ to $b$ should be dominated by the KL-divergence. 
The inequality \eqref{non-gauss-ineq} expresses this expectation in quantitative form. 
The change in the expectation of $r$ is bounded by a power series in the KL divergence, whose coefficients are the cumulants of $r$ in the reference system $a$. 
Thus, the inequality \eqref{non-gauss-ineq} establishes a universal relation between the response of the observable $r$ to the perturbation and the fluctuations of $r$ in the reference system. 
\eqref{non-gauss-ineq} is our first main result and the most general form of our fluctuation-response inequality (FRI).
We remark that, even when the infimum on the right-hand side of \eqref{non-gauss-ineq} cannot be evaluated explicitly, we still obtain an upper bound on the response by choosing an arbitrary value of $\alpha$, e.~g.~$\alpha = 1$ corresponding to the Gaussian case.

\subsection{Linear response}
While the FRI \eqref{non-gauss-ineq} is valid for arbitrary systems $a$ and $b$, and thus arbitrarily strong perturbations, the right-hand side involves all cumulants of the observable $r$ in the reference system. 
In many cases, the high-order cumulants are difficult to evaluate either from a theoretical model or from a measurement. 
However, if the perturbation is weak, such that the probability density of system $b$ differs from the one in system $a$ only by terms of order $\epsilon$ with $\epsilon \ll 1$, then the KL divergence is of order $\epsilon^2$. 
We can thus neglect the higher-order terms in \eqref{non-gauss-ineq} and obtain our second main result, the FRI for linear response (LFRI)
\begin{align}
\big\vert \av{r}^b - \av{r}^a \big\vert \leq \sqrt{2 D_{\text{KL}}^{b \Vert a} \av{\Delta r^2}^a} + O(\epsilon^{\frac{3}{2}}) \label{FRI}.
\end{align}
We may use the condition $D_\text{KL}^{b \Vert a} = O(\epsilon^2)$ as an information-theoretic definition of the linear response regime. 
Whenever this condition is satisfied, \eqref{FRI} guarantees that the change in any observable is at most of order $\epsilon$. 
In the linear response regime, the response of any observable to the perturbation is thus bounded from above by the variance of the observable in the reference system times the KL divergence between the perturbed and unperturbed probability densities. 
We remark that \eqref{FRI} is exactly the same as \eqref{gauss-ineq}. 
Thus the linear response result \eqref{FRI} remains valid beyond linear response, provided that the distribution of $r$ in the reference system is Gaussian.

\subsection{Physical interpretation of KL-divergence}
In general, the KL divergence \eqref{KL-divergence} is an information-theoretic quantity and its calculation requires knowledge of the explicit distributions $P^a(\omega)$ and $P^b(\omega)$. 
However, in specific situations, the KL divergence can be expressed in terms of physical observables. 
One such situation is when $P^a(\omega)$ and $P^b(\omega)$ represent the path probabilities of continuous-time Markov processes, for example a Markov jump or diffusion process. 
In the former case, we consider a set of $N$ discrete states, with jumps from state $j$ to state $i$ occurring at a rate $W^a_{i j}(t)$. 
In terms of the rates, the probability to jump from state $j$ to state $i$ in the infinitesimal time interval $[t,t+dt]$ is given by $W^a_{i j}(t) dt$. 
The probability $p^a_i(t)$ to be in state $i$ at time $t$ then evolves according to the Master equation
\begin{align}
d_t p^a_i(t) = \sum_j \big( W^a_{i j}(t) p^a_j(t) - W^a_{j i}(t) p^a_i(t) \big)  \label{master}
\end{align}
with prescribed initial probabilities $p^a_i(0)$. 
We also consider a second Markov jump process on the same state space, however, with transition rates $W_{i j}^b(t)$ and initial probabilities $p_i^b(0)$. 
The linear response requirement that the two Markov jump processes should be close to each other is realized by choosing
\begin{align}
W^b_{i j}(t) = W^a_{i j}(t) e^{\epsilon \mathcal{Z}_{i j}(t)} \quad \text{and} \quad p_i^b(0) = p_i^a(0) \big(1 + \epsilon q_i \big) 
\end{align}
with $\epsilon \ll 1$ and $\mathcal{Z}_{i i}(t) = 0$. 
The $\mathcal{Z}_{i j}(t)$ and $q_i$ are parameters of at most order $1$, which characterize the difference between the two jump processes. 
As we show in the Supporting Information, the KL divergence between the probabilities of observing a given trajectory in the respective jump processes during a time interval $[0,\mathcal{T}]$ can be expressed as
\begin{align}
D_\text{KL}^{b \Vert a} = \frac{\epsilon^2}{2} \bigg( \int_0^\mathcal{T} dt \sum_{i,j} \mathcal{Z}_{i j}(t)^2 W^a_{i j}(t) p^a_j(t) + \sum_{i} (q_i)^2 p_i^a(0) \bigg) \label{KL-markov} 
\end{align}
where we neglect terms of order $\epsilon^3$ and higher. 
As a more concrete example, we consider a parameterization of the transition rates as
\begin{align}
W_{i j}^a = k_{i j} e^{-\frac{\beta}{2} \big( E_i - E_j + B_{i j} - \sum_\nu \mathcal{A}_{i j}^\nu f^\nu(t) \big)} \label{rate-param},
\end{align}
where $E_i$ is the energy of state $i$, $B_{i j} = B_{j i}$ is an energy barrier separating the states $i$ and $j$ and $\beta = 1/(\kb T)$ is the inverse temperature. 
We further set $k_{i j} = k_{j i} = 1$ if a transition between two states is possible and zero otherwise. $f^\nu(t)$ represent a set of generalized forces that drive the system out of equilibrium, with the coefficients $\mathcal{A}_{i j}^\nu = - A_{j i}^\nu$ determining how the force $\nu$ impacts the transition rates. 
For $f^\nu \equiv 0$ for all $\nu$, the steady state of the system is the Boltzmann equilibrium $p_i^\text{eq} \sim e^{-\beta E_i}$, while for $f^\nu \neq 0$ the system is out of equilibrium.
Note that the parameterization \eqref{rate-param}, with appropriate choices for the parameters covers all possible cases of \eqref{master} which satisfy $W_{i j} = 0 \Leftrightarrow W_{j i} = 0$.
As a perturbation, we change the generalized forces to $f^\nu(t) + \epsilon \phi^\nu(t)$ while keeping the initial state fixed. This yields
\begin{align}
D_\text{KL}^{b \Vert a} = \frac{\epsilon^2}{2} \int_0^\mathcal{T} dt \sum_{\mu,\nu} \phi^\mu(t) \mathcal{L}^{\mu \nu}(t) \phi^\nu(t),
\end{align}
where we defined
\begin{align}
\mathcal{L}^{\mu \nu}(t) = \frac{\beta^2}{4} \sum_{i,j} \mathcal{A}_{i j}^\mu \mathcal{A}_{i j}^\nu W_{i j}^a(t) p_j^a(t) .
\end{align}
In particular, the KL divergence between the path probabilities is determined by the change in the generalized forces and thus can be measured or calculated explicitly.

In the second case, a diffusion process, we have a set $\bm{x}(t) = (x_1(t),\ldots,x_N(t))$ of continuous stochastic variables, whose time evolution is given by the Langevin equation
\begin{align}
\dot{\bm{x}}(t) = \bm{a}(\bm{x}(t),t) + \sqrt{2 \bm{B}(\bm{x}(t),t)} \cdot \bm{\xi}(t) \label{langevin} ,
\end{align}
where the drift vector $\bm{a}(\bm{x},t)$ contains the systematic generalized forces and the positive definite and symmetric diffusion matrix $\bm{B}(\bm{x},t)$ describes how the system is coupled to a set of mutually independent Gaussian white noises $\av{\xi_i(t) \xi_j(s)} = \delta_{ij} \delta(t-s)$. 
Here, the symbol $\cdot$ denotes the It{\=o}-product and we specify the initial probability density $p^a(\bm{x},0)$.
As in the Markov jump case, we consider a second process with a different drift vector $\bm{b}(\bm{x},t)$ and initial probability density $p^b(\bm{x},0)$. 
Importantly, in order to ensure a finite relative entropy between the path probabilities, the diffusion matrix of both processes has to be the same. 
We ensure that the process $b$ is related to $a$ in linear response by setting
\begin{align}
\bm{b}(\bm{x},t) &= \bm{a}(\bm{x},t) + \epsilon \bm{\alpha}(\bm{x},t) \label{modification-diff} \\
 \text{and} \quad p^b(\bm{x},0) &= p^a(\bm{x},0)\big(1+\epsilon q(\bm{x}) \big) \n  .
\end{align}
As show in the Supporting Information, in this case, the KL divergence between the path probabilities evaluates to
\begin{align}
D_\text{KL}^{b \Vert a} &= \frac{\epsilon^2}{2} \bigg( \frac{1}{2} \int_0^\mathcal{T} dt \int d\bm{x} \ \bm{\alpha}(\bm{x},t)^T \bm{B}(\bm{x},t)^{-1} \bm{\alpha}(\bm{x},t) p^a(\bm{x},t) \nonumber \\
& \hspace{1.5 cm} +  \int d\bm{x} \ \big(q(\bm{x})\big)^2 p^a(\bm{x},0) \bigg) \label{KL-diffusion},
\end{align}
where the superscript $T$ denotes transposition, and again neglecting higher-order terms in $\epsilon$. Here, the identification of the KL divergence with the change in the generalized forces is immediate.

Both for a Markov jump process and a diffusion process, the KL divergence between the path probabilities between the perturbed and reference system thus can be expressed as an observable, whose average is evaluated in the reference system. 
This gives the LFRI \eqref{FRI} additional physical meaning, as it provides a universal relation between different physical observables: 
On the one hand, we have the observable $r$, its response to the perturbation and its fluctuations. 
On the other hand, we have the KL divergence expressed in terms of the magnitude of the perturbing generalized forces. 
Since the KL divergence is evaluated for the path probabilities of the process, we may choose any observable (including observables measured at a given time, correlation functions and stochastic currents) that depends on the path and \eqref{FRI} remains valid.
We stress that both the reference system and the perturbation are in principle arbitrary. 
The reference system is not restricted to equilibrium or even steady states, but we may also consider for example time-dependent perturbations of an already time-dependent reference system. 
Also, while the perturbation may represent a physical force, we may also choose a more general type of perturbation, which does not represent any force realizable in practice, but for which the KL divergence has a physical interpretation. 
We will discuss examples of both kinds of perturbation in the next sections.

\section{Mobility and diffusion}
As a direct application of the LFRI, we consider a diffusion process of $N$ particles in contact with a heat bath. We take $\bm{a} = \bm{M} \bm{f}$, $\bm{B} = \kb T \bm{M}$ \cite{Lau07,Mae08,Far14} in \eqref{langevin}, where $\bm{f}(\bm{x},t)$ is a force, $\bm{M}$ is the mobility matrix and $T$ the temperature of the heat bath,
\begin{align}
\dot{\bm{x}}(t) = \bm{M} \bm{f}(\bm{x}(t),t) + \sqrt{2 \kb T \bm{M}} \bm{\xi}(t) .
\end{align}
The force $\bm{f}(\bm{x},t)$ may contain global potential forces, interactions between the particles and also non-conservative driving forces. We assume that for long times, the system exhibits an asymptotic drift velocity $\bm{v}^\text{d}$,
\begin{align}
\av{\bm{x}(\mathcal{T}) - \bm{x}(0)} \simeq \mathcal{T} \bm{v}^\text{d}.
\end{align}
This may be realized for example for particles diffusing in a periodic potential under the influence of a an external force or time-periodic driving.
We now apply an additional small constant force $\bm{\varphi}$ to the system, i.~e.~consider the system with total force $\bm{f}+\bm{\varphi}$ as the perturbed system.
In general, this will change the drift velocity to $\tilde{\bm{v}}^{\text{d}}$.
We then define the differential mobility matrix $\bm{\mathcal{M}}$ via
\begin{align}
\tilde{\bm{v}}^{\text{d}} - \bm{v}^\text{d} =  \bm{\mathcal{M}} \bm{\varphi} \label{effective-mobility},
\end{align}
or, component-wise $\mathcal{M}_{i j} = \partial v^\text{d}_i/\partial \varphi_j \vert_{\varphi = 0}$.
In the absence of the forces $\bm{f}$, the differential mobility is just equal to the bare mobility $\bm{\mathcal{M}} = \bm{M}$.
However, if $\bm{f}$ is non-zero, e.~g.~if the particles move in a periodic potential or interact with each other, the expression for the differential mobility quickly becomes more complicated and can in general no longer be evaluated explicitly.
For the above perturbation, the KL divergence \eqref{KL-diffusion} is given by
\begin{align}
D_\text{KL}^{b \Vert a} \simeq \frac{\mathcal{T}}{4 \kb T} \bm{\varphi}^T \bm{M} \bm{\varphi} .
\end{align}
As the observable $r$, we choose the projection of $\bm{x}$ on some constant vector $\bm{e}$, $r = \bm{e} \bm{x}$.
Defining the diffusivities $D_{i j} = \lim_{\mathcal{T} \rightarrow \infty} \av{\Delta x_i \Delta x_j}_\mathcal{T}/(2\mathcal{T})$ and using the definition of the differential mobility \eqref{effective-mobility}, we then get from the LFRI \eqref{FRI} for long times
\begin{align}
\big( \bm{e} \bm{\mathcal{M}} \bm{\varphi} \big)^2 \leq \frac{1}{\kb T}  \big(\bm{e} \bm{D} \bm{e}\big) \ \big( \bm{\varphi} \bm{M} \bm{\varphi} \big) .
\end{align}
Since $\bm{e}$ and $\bm{\varphi}$ are arbitrary, we find the bound on any component of the differential mobility tensor
\begin{align}
\big(\mathcal{M}_{i j}\big)^2 \leq \frac{D_{i i} M_{j j}}{\kb T} \label{mobility-bound} .
\end{align}
This inequality is our third main result and imposes a strong constraint on particle transport:
The mobility in direction $i$ in response to a force in direction $j$ is bounded by the diffusion coefficient in direction $i$ times the bare mobility in direction $j$.
For a particle in a tilted, one-dimensional periodic potential, the equivalent of \eqref{mobility-bound} was derived in Ref.~\cite{Hay05}.
In particular we have for the diagonal components $(\mathcal{M}_{ii})^2 \leq (M_{ii})^2 D_{ii}/D_{ii}^\text{free}$, where $D_{ii}^\text{free} = \kb T M_{ii}$ is the free-space diffusivity in the absence of any force. 
Thus, enhancing the mobility beyond its bare value necessarily requires enhanced diffusivity.
In equilibrium, there is a one-to-one correspondence between the diffusion coefficient and the mobility in the form of the fluctuation-dissipation theorem $D_{i j}^\text{eq} = \kb T \mathcal{M}_{i j}^\text{eq}$.
This simple relation breaks down in out-of-equilibrium situations \cite{Mae08}.
It is, however, useful to define effective temperatures in analogy to the equilibrium case via $D_{ii} = \kb T_i^\text{eff} \mathcal{M}_{ii}$ \cite{Cug97,Hay04}.
The bound \eqref{mobility-bound} then translates into a lower bound for the effective temperatures
\begin{align}
T_i^\text{eff} \geq \frac{\mathcal{M}_{ii}}{M_{ii}} T \label{temperature-bound} .
\end{align}
Since we have $T = T_i^\text{eff}$ in an equilibrium system, the bound \eqref{temperature-bound} tells us that we must have $\mathcal{M}_{ii}^\text{eq} \leq M_{ii}$, i.~e.~enhancing the mobility beyond its free-space value is only possible in a non-equilibrium situation.
For non-equilibrium systems with enhanced mobility (e.~g.~a periodic potential close to critical tilt \cite{Rei01}), the effective temperature is always larger than the physical temperature.
On the other hand, if the effective temperature is lower than the physical one \cite{Sas05,Nak13}, \eqref{temperature-bound} predicts that the mobility is reduced by at least a factor $T^\text{eff}/T$.
We remark that the bound \eqref{mobility-bound} applies to equilibrium systems, non-equilibrium steady states, situations where the potential varies periodically in time or fluctuates, and also to underdamped dynamics.

\section{Symmetries and thermodynamic inequalities}
Another example in which the KL divergence \eqref{KL-divergence} acquires a physical meaning is when considering a stochastic process and its time reverse.
Specifically, we choose $P^b(\omega)$ to be the path probability density $\mathbb{P}(\omega)$ of a stochastic process and $P^a(\omega)$ as the probability density of the time-reversed trajectory $\omega^\dagger$ under an appropriate time-reversed dynamics with path probability $\mathbb{P}^\dagger(\omega)$, see Refs.~\cite{Sei12,Spi12,Gar12} for a more detailed discussion.
In this case, the KL divergence is equal to the total entropy production,
\begin{align}
D_\text{KL}^{b \Vert a} = \Delta S = \int d\omega \ \mathbb{P}(\omega) \ln \bigg( \frac{\mathbb{P}(\omega)}{\mathbb{P}^\dagger(\omega^\dagger)} \bigg) \label{KL-entropy}.
\end{align}
From the FRI \eqref{non-gauss-ineq} we then immediately obtain
\begin{align}
&\big\vert \av{r} - \av{r}^\dagger \big\vert \leq \sqrt{2 \Delta S \av{\Delta r^2}^\dagger}  \\
& \quad \times \inf_{\alpha > 0} \Bigg[ \frac{1}{2 \alpha} + \frac{1}{2 \alpha \Delta S} K_{\Delta r}^\dagger \Bigg( \sqrt{\frac{2 \Delta S}{\av{\Delta r^2}^\dagger}} \sigma \alpha \Bigg) \Bigg] \n ,
\end{align}
where the superscript $\dagger$ denotes that the average is taken with respect to the time-reversed path probability density $\mathbb{P}^\dagger(\omega^\dagger)$.
Thus the change in any observable under time-reversal is dominated by the entropy production and the fluctuations of the observable in the time-reversed dynamics.
In particular, if the observable is odd under time-reversal, such that $\kappa_{n,r}^\dagger = (-1)^n \kappa_{n,r}$, then the above simplifies to
\begin{align}
&\sqrt{2} \av{r}  \leq \sqrt{\Delta S \av{\Delta r^2}} \label{non-linear-TUR}  \\
& \quad \times \inf_{\alpha > 0} \Bigg[ \frac{1}{2 \alpha} + \frac{1}{2  \alpha \Delta S} K_{\Delta r} \Bigg(- \sqrt{\frac{2 \Delta S}{\av{\Delta r^2}}} \alpha \Bigg) \Bigg] \n ,
\end{align}
where we assumed $\av{r} \geq 0$ without loss of generality.
We note that the first factor on the right-hand side is reminiscent of the thermodynamic uncertainty relation (TUR) \cite{Bar15,Gin16}
\begin{align}
2\big(\av{r}\big)^2  \leq \Delta S \av{\Delta r^2} \label{TUR},
\end{align}
which holds for time-integrated currents $r$ in the steady state of a continuous-time Markovian dynamics.
\eqref{non-linear-TUR} provides a generalization of the TUR to arbitrary dynamics and observables, as long as the latter are odd under time reversal.
The trade-off is that, without specifying the dynamics, we generally need to take into account higher-order cumulants of the fluctuations.
From this point of view, a natural question is why the TUR \eqref{TUR} does not involve higher-order cumulants, even though the distribution of the current is generally non-Gaussian.
The answer is that \eqref{non-linear-TUR} arises as a consequence of the symmetry of the observable under a discrete transformation, namely time-reversal.
For this discrete transformation, the KL divergence in the FRI \eqref{non-gauss-ineq} is generally not small and we cannot neglect the higher-order terms.
On the other hand, as we will show in the following, \eqref{TUR} is actually the consequence of a continuous symmetry, which allows us to consider an infinitesimal transformation for which the KL divergence in \eqref{non-gauss-ineq} is small and we thus can use the LFRI \eqref{FRI}.

We remark that, from \eqref{non-linear-TUR}, we can infer that the TUR always holds irrespective of the dynamics if the distribution of the observable is Gaussian.
This covers, for example, linear but non-Markovian generalized Langevin dynamics \cite{Spe07}.
Further, while \eqref{non-linear-TUR} involves higher powers of the entropy production, in contrast to another recently derived bound, which is exponential in the entropy production \cite{Has19},
\begin{align}
\sqrt{2} \av{r} \leq \sqrt{(e^{\Delta S} - 1) \av{\Delta r^2}} \label{exp-bound},
\end{align} 
it remains useful in the long-time or large-system limit.
For non-equilibrium systems with short-range interactions and finite correlation time, both the entropy production and the cumulants of a stochastic current are generally extensive in both time time and system size \cite{Nem11}.
Consequently, while the bound \eqref{exp-bound} grows exponentially in time and system size, the term in square brackets in \eqref{non-linear-TUR} remains of $1$ in theses limits.
Since \eqref{non-linear-TUR} and \eqref{exp-bound} both only require the observable to be odd under time-reversal, we anticipate \eqref{non-linear-TUR} to be more useful when applied to macroscopic systems at finite time.

\subsection{Steady-state thermodynamic uncertainty relation}
We now want to derive the steady-state TUR \eqref{TUR} from the LFRI \eqref{FRI}.
We specialize the discussion to a diffusion process \eqref{langevin}, which we describe in terms of its probability density $p^a(\bm{x},t)$, which obeys the Fokker-Planck equation
\begin{align}
\partial_t p^a(\bm{x},t) &= - \bm{\nabla}^T \bm{j}^a(\bm{x},t) \qquad \text{with} \label{fokkerplanck}\\
\bm{j}^a(\bm{x},t) &= \big(\bm{a}(\bm{x},t) - \bm{\nabla}^T \bm{B}(\bm{x},t) \big) p(\bm{x},t) \n,
\end{align}
where $\bm{\nabla}$ is the gradient with respect to $\bm{x}$.
For general diffusion dynamics of the type \eqref{fokkerplanck}, the total entropy production in the system and its environment is given by \cite{Spi12}
\begin{align}
\Delta S &= \int_0^\mathcal{T} dt \int d\bm{x} \ \frac{(\bm{j}^{a,\text{irr}})^T \bm{B}^{-1} \bm{j}^{a,\text{irr}}}{p^a} \label{entropy}  ,
\end{align}
where $\bm{j}^{a,\text{irr}}$ is the irreversible probability current, corresponding to forces that are odd under time-reversal.
We now choose the additional force in \eqref{modification-diff} as
\begin{align}
\bm{\alpha}(\bm{x},t) &= \frac{\bm{j}^{a,\text{irr}}(\bm{x},t)}{p^a(\bm{x},t)} \label{entropy-mod}.
\end{align}
This perturbation does not generally correspond to any physically realizable force; we thus refer to it as a \enquote{virtual} perturbation.
For this choice, it is obvious that the KL divergence \eqref{KL-diffusion} corresponds to the entropy production \eqref{entropy}.
In case of a steady state dynamics that is even under time reversal, $\bm{j}^a = \bm{j}^{a,\text{irr}}$, the effect of the perturbation \eqref{entropy-mod} is simply a rescaling of the steady-state probability currents, $\bm{j}^b(\bm{x}) = (1+\epsilon) \bm{j}^a(\bm{x})$.
Choosing a time-integrated generalized current as the observable \cite{Che15}, whose average is given by
\begin{align}
\av{r}^a = \int_0^\mathcal{T} dt \int d\bm{x} \ \bm{\chi}(\bm{x},t) \bm{j}^a(\bm{x},t),
\end{align}
with some vector-valued function $\bm{\chi}$, we then have $\av{r}^{b} = (1+\epsilon) \av{r}^a$ and thus
\begin{align}
2\big(\av{r}\big)^2 \leq \av{\Delta r^2} \Delta S \label{uncertainty} ,
\end{align}
where we suppressed the explicit reference to the dynamics $a$.
This is precisely the finite-time uncertainty relation \eqref{TUR} proposed in Ref.~\cite{Pie17} and proven in Refs.~\cite{Hor17,Dec17}, for jump and diffusion processes, respectively.
Surprisingly, comparing the dynamics of the reference system $a$ to the dynamics of a system $b$ with an appropriately tailored \enquote{virtual} perturbation, can reveal properties of the reference system itself; in this case, that it obeys the thermodynamic uncertainty relation.

In most physical situations, the irreversible probability currents describe the interaction between the system and the surrounding heat bath. 
The perturbation \eqref{entropy-mod} thus corresponds to slightly changing the strength of this interaction.
If the entire dynamics of the system is driven by the heat bath, the coupling to the heat bath sets the overall time scale of the dynamics and the magnitude of the probability currents.
By contrast, if other timescales not governed by the heat bath (e.~g.~due to time-dependent forces or reversible dynamics) are present in the system, then the average current $\av{r}$ will generally not satisfy an uncertainty relation of the type \eqref{uncertainty}.

\subsection{Demonstration: Markov chain model}
\begin{figure}
\includegraphics[width=.48\textwidth]{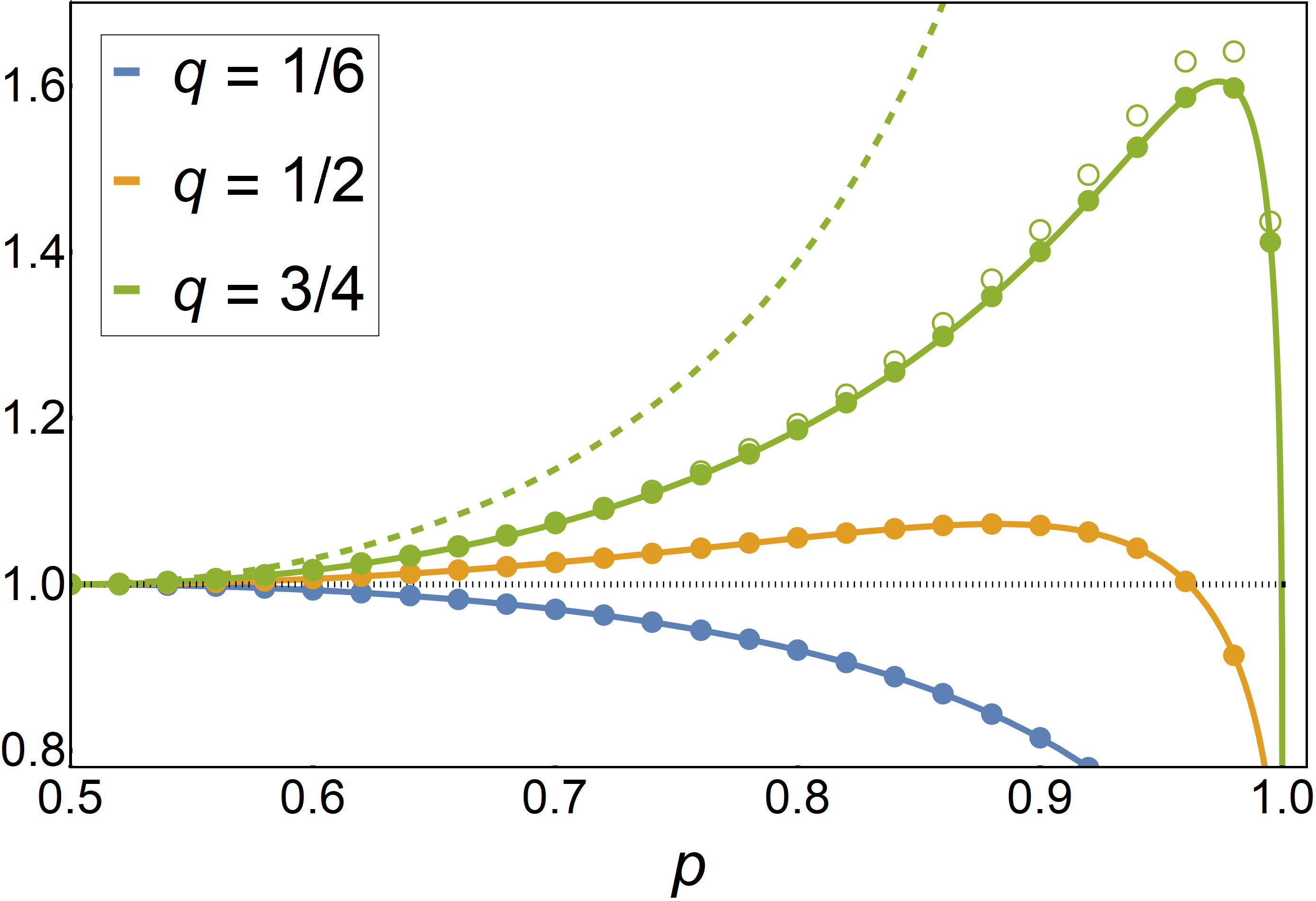}
\caption{The ratio $2 \av{r}^2/(\av{\Delta r^2} \Delta S)$ for the Markov-chain model \eqref{transition-prob} as a function of the asymmetry $p$ for various jump probabilities $q$. The solid lines are the exact results \eqref{av-var-markov-chain} and \eqref{entropy-markov-chain}, the dotted line is the TUR bound \eqref{TUR}. For sufficiently large jump probability (orange, green) the ratio can exceed the TUR.
By contrast, the bound \eqref{non-linear-TUR} (filled symbols) holds and in this case yields an equality.
For comparison, we also show the bound obtained by evaluating \eqref{non-linear-TUR} at the Gaussian value of $\alpha = 1$ (empty symbols) and the exponential bound obtained in Refs.~\cite{Pro17,Has19} (dashed line) for the case $q = 3/4$.}\label{fig-ratio}
\end{figure}

One important class of dynamics, which, in general, does not satisfy the TUR \eqref{TUR} even in the long-time limit, is a discrete-time Markov chain.
Similar to \eqref{master}, we consider a set of $N$ states, however, the transitions between the states do not follow a rate process, but instead occur only at discrete times $t = 1, \ldots, M$.
Consequently, the transition rates $W_{i j}$ are replaced by transition probabilities $Q_{i j}$ from state $j$ to state $i$,
\begin{align}
p_i^{t} = \sum_{j} Q_{i j} p_j^{t-1} \label{markov-chain},
\end{align}
where $p_i^t$ is the probability to be in state $i$ at step $t$, and the transition probabilities satisfy the normalization condition $\sum_i Q_{i j} = 1$.
As a concrete example, we consider a discrete-time random walk on a ring of $N$ sites.
The random walker jumps with probabilities $q p$ and $q (1-p)$ to from site $i$ to $i+1$ and from $i$ to $i-1$, respectively, where $p,q \in [0,1]$.
The probability to remain at site $i$ is $1-q$.
The transition probabilities are then given by
\begin{align}
Q_{i,i} = 1-q, \quad Q_{i+1,i} = q p, \quad Q_{i,i-1} = q (1-p) \label{transition-prob},
\end{align}
where we identify $N+1 \leftrightarrow 1$ and $0 \leftrightarrow N$.
This model has also be investigated for $N = 3$ in Ref.~\cite{Rol19}.
We assume that the initial occupation probabilities are $p_i^0 = 1/N$, which is the steady state of \eqref{markov-chain}.
We define the observable $r$ such that it increases by $1/N$ whenever the walker jumps from $i$ to $i+1$ and decreases by $1/N$ upon a jump from $i$ to $i-1$.
Thus, $r$ counts the number of times the walker has gone around the ring in positive direction.
For this observable, it is easy to see that it is odd under time-reversal.
We further can compute the cumulant generating function explicitly,
\begin{align}
K_r(h) = M \ln \Big( 1 - q + e^{\frac{h}{N}} q p - e^{-\frac{h}{N}} q (1-p) \Big),
\end{align}
where $M$ is the number of steps.
In this case, the linear scaling of the cumulant generating function and thus the cumulants with the number of steps is explicit.
Using these results, it is straightforward to obtain
\begin{align}
\av{r} = \frac{M}{N} q (1-2 p), \quad \av{\Delta r^2} = \frac{M}{N^2} q (1- q(1-2 p)^2) \label{av-var-markov-chain} .
\end{align}
We further have for the entropy production,
\begin{align}
\Delta S = M \Bigg( q p \ln\bigg(\frac{p}{1-p}\bigg) + q (1-p) \ln\bigg(\frac{1-p}{p}\bigg) \Bigg) \label{entropy-markov-chain} .
\end{align}
In Fig.~\ref{fig-ratio}, we plot the ratio $2 \av{r}^2/(\av{\Delta r^2} \Delta S)$ as a function of the bias $p$ for various jump probabilities $q$.
For a system satisfying the TUR \eqref{TUR}, this ratio is bounded by $1$ (dotted line).
It can be seen that, while for small jump probability, the TUR is satisfied, the dynamics can violate the TUR for $q > 1/3$.
This hints at the reason why the TUR holds for continuous-time dynamics but can be violated for a discrete-time one:
In the continuous-time case, the probability to observe no transition at all (i.~e.~the staying probability) is close to $1$ for short times.
By contrast, the staying probability (in the present example $1-q$) for a single step can be significantly less that $1$ in the discrete-time case.
This prevents us from finding an infinitesimal transformation, which rescales the current and whose KL-divergence is equal to the entropy production.
While the TUR can thus be violated for discrete-time dynamics, the bound \eqref{non-linear-TUR}, which makes no assumptions on the nature of the dynamics, holds.

In this specific case, we actually obtain equality in \eqref{non-linear-TUR}, showing that the bound is tight.
To understand this, we note that we have from \eqref{cgf-bound} and \eqref{KL-entropy}
\begin{align}
\Delta S \geq \sup_{h} \big( h \av{r} - K_r(-h) \big) \label{entropy-identity},
\end{align}
and attaining equality is equivalent to equality in \eqref{non-linear-TUR}.
In the present case, the ratio between the path probabilities can simply be written as
\begin{align}
\ln \bigg( \frac{\mathbb{P}(\omega)}{\mathbb{P}^\dagger(\omega^\dagger)} \bigg) = \sum_{t = 1}^{M} \big( \delta_{t +} - \delta_{t -} \big) \ln\bigg(\frac{p}{1-p}\bigg) + \ln\bigg(\frac{p_{i_0}^0}{p_{i_0}^M}\bigg) ,
\end{align}
where $\delta_{t+}$ ($\delta_{t-}$) is $1$ if a jump from $i$ to $i+1$ ($i$ to $i-1$) occurs in the step $t-1 \rightarrow t$ and zero otherwise.
The last term vanishes in the steady state, and the first term is just equal to $N \ln(p/(1-p)) r(\omega)$, where $r(\omega)$ is the cycle-counting observable defined above.
Thus, the observable $r$ is up to a constant factor equal to the logarithmic ratio of the forward and reverse path probability.
In other words, the observable $r$ contains all the information about the irreversibility of the dynamics.
For the choice $h = N \ln(p/(1-p))$, we have equality in \eqref{entropy-identity} with our observable $r$ and the corresponding choice of $\alpha$ leads to equality in \eqref{non-linear-TUR}.
%Cast in the language of the derivation of the TUR in the previous section, this means that we can no longer find an infinitesimal transformation that leads to a rescaling of the current: any such transformation necessarily modifies both the transition and staying probabilities since their sum has to be conserved.

\section{Discussion}
Information-theoretic concepts have mostly been employed in thermodynamics to account for Maxwell's demon and feedback  \cite{Sag10,Esp11,Bar14,Hor14}.
The LFRI \eqref{FRI} establishes a universal relation between the linear response of an observable, its fluctuations and the KL divergence for arbitrary non-equilibrium states, evidencing that also the classical topic of linear response can be understood in terms of information.
The KL divergence characterizes the information about the perturbation contained in the respective probability densities.
This bounds the amount of information contained in any observable and thus puts a limit on the magnitude of the response.

The mathematical basis of the FRI is the bound \eqref{relent-bound}.
Similar bounds have previously been used in the context of large deviation theory \cite{Hol08,Che15B}, where the goal is to find an optimal tilted process $P^b$ that turns the bound into an equality.
The present discussion shows that the bound remains useful when replacing the tilted process by a physical, perturbed process, yielding relations between physical observables.

The motivation of response theory is to infer the properties of a physical system by observing its response to perturbations.
The FRI provides a new, intriguing way of applying this principle to obtain universal relations between observables.
Instead of measuring the effect of a specific perturbation on the system, we tailor the perturbation such that the KL-divergence can be expressed as a physical observable.
For such a \enquote{virtual} perturbation, the FRI then yields an inequality between physical observables.
In this article, we outlined two such applications of the FRI: a relation between mobility and diffusivity in particle transport and the re-derivation of the thermodynamic uncertainty relation and its extension to arbitrary dynamics.
Since the FRI holds for a wide range of dynamics, observables and perturbations, we anticipate that it can serve as a starting point for the derivation of many more relations between physical quantities.

\begin{acknowledgments}
\textbf{Acknowledgments.}
This work was supported by KAKENHI (Nos. 17H01148, 19H05496, 19H05795), World Premier International Research Center Initiative (WPI), MEXT, Japan and by JSPS Grant-in-Aid for Scientific Research on Innovative Areas "Discrete Geometric Analysis for Materials Design": Grant Number 17H06460.
The authors wish to thank R. Chetrite, S. Ito and T. Sagawa for helpful discussions.
\end{acknowledgments}

\appendix

\onecolumngrid 

\section{Kullback-Leibler divergence for Markov jump and diffusion processes}

In the following, we derive explicit expressions for the Kullback-Leibler divergence between the path probabilities (probability densities) of two Markov jump, respectively diffusion processes.
In full generality, we here treat the case of a combined jump-diffusion process, which is governed by the Fokker-Planck equations,
\begin{align}
\partial_t p_k(\bm{x},t) &= - \bm{\nabla} \bm{j}^\text{d}_{k}(\bm{x},t)  + \sum_{l = 1}^{N^\text{j}} j^\text{j}_{k l}(\bm{x},t) \label{sup-fpe} \\
\text{with} \quad \bm{j}^\text{d}_{k}(\bm{x},t) = \Big(\bm{a}_k(\bm{x},t) - \bm{\nabla} \bm{B}_k(\bm{x},t) \Big) p_k(\bm{x},t) &\qquad
\text{and} \qquad j^\text{j}_{k l}(\bm{x},t) = W_{k l}(\bm{x},t) p_l(\bm{x},t) - W_{l k}(\bm{x},t) p_k(\bm{x},t) \n,
\end{align}
with given initial density $p_k(\bm{x},0)$.
Here $\bm{x}(t) = (x_1(t),\ldots,x_{N^d}(t))$ denotes a collection of $N^\text{d}$ continuous degrees of freedom which we refer to as space, $\bm{a}_k(\bm{x},t)$ is a drift vector and $\bm{B}_k(\bm{x},t)$ a symmetric, positive definite diffusion matrix, both of which may depend on the continuous variables $\bm{x}$, on time $t$ and also on the state variable $k$, whose dynamics we assume to be governed by a Markov jump process with time- and space-dependent transition rate $W_{k l}(\bm{x},t)$ from state $l$ to state $k$.
An explicit example for such a combined process is a flashing ratchet, i.~e.~a particle diffusing under the influence of a potential which can randomly switch between two different functional forms.
In \eqref{sup-fpe}, we also defined the diffusive probability currents $\bm{j}^\text{d}$ and the jump probability currents $j^\text{j}$.
For $j^\text{j} = 0$ or $\bm{j}^\text{d}=0$, a pure diffusion, respectively jump, process is recovered.
We start by constructing the short-time propagator $p(\bm{x},k,t+\tau \vert \bm{y},m,t)$ for being in state $(\bm{x},k)$ at time $t+\tau$ starting from a state $(\bm{y},m)$ at time $t$.
Obviously, we have $p(\bm{x},k,t \vert \bm{y},m,t) = \delta(\bm{x}-\bm{y}) \delta_{k m}$.
For sufficiently small $\tau$, we can use \eqref{sup-fpe} to obtain the approximate transition probability to first order in $\tau$,
\begin{align}
p(\bm{x},k,t+\tau \vert \bm{y},m,t) = \delta(\bm{x}-\bm{y}) \delta_{k m} - &\tau \bm{\nabla} \Big(\bm{a}_m(\bm{y},t) - \bm{\nabla} \bm{B}_m(\bm{y},t) \Big) \delta(\bm{x}-\bm{y}) \delta_{k m} \label{prop-jd-1} \\
& + \tau \Big( W_{k m}(\bm{y},t) - \delta_{k m} \sum_l W_{m l}(\bm{y},t) \Big) \delta(\bm{x}-\bm{y}) + O(\tau^2) \n .
\end{align}
The first two terms can to leading order be replaced by the Gaussian propagator of the diffusion process $p^{(m)}$ in the (fixed) state $m$,
\begin{align}
p^{(m)}(\bm{x},t+\tau &\vert \bm{y},t) = \delta(\bm{x}-\bm{y})  - \tau \bm{\nabla} \Big(\bm{a}_m(\bm{y},t) - \bm{\nabla} \bm{B}_m(\bm{y},t) \Big) \delta(\bm{x}-\bm{y}) + O(\tau^2) \\
&= \frac{1}{\sqrt{4 \pi \det(\bm{B}_m(\bm{y},t)) \tau }} \exp \bigg[ - \frac{1}{4 \tau} \Big( \bm{x} - \bm{y} - \bm{a}_m(\bm{y},t) \tau \Big)^T \big(\bm{B}_m\big)^{-1}(\bm{y},t) \Big(\bm{x} - \bm{y} - \bm{a}_m(\bm{y},t) \tau \Big) \bigg] + O(\tau^2) \n .
\end{align}
Since the third term on the right-hand side of \eqref{prop-jd-1} is already of order $\tau$, we can also replace the delta-function by the propagator of the diffusion process and obtain for the propagator of the combined jump-diffusion process
\begin{align}
p(\bm{x},k,t+\tau \vert \bm{y},m,t) &= p^{(m)}(\bm{x},t+\tau \vert \bm{y},t) p^{(\bm{y})}(k,t+\tau \vert m,t) \label{prop-jd} \\
\text{with} \qquad p^{(\bm{y})}(k,t+\tau \vert m,t) &= \delta_{k m} + \tau \Big( W_{k m}(\bm{y},t) - \delta_{k m} \sum_l W_{m l}(\bm{y},t) \Big) + O(\tau^2) \n .
\end{align}
To leading order in $\tau$, we can thus write the propagator of the jump-diffusion process as a product of diffusion and jump propagators.
This represents the fact that the probability of observing a jump and significant spatial motion at the same time is of order $\tau^2$ for sufficiently small $\tau$.
This factorization carries over to the path probability density, which is expressed as a product of transition probabilities,
\begin{align}
\mathbb{P} = \prod_{i = 1}^N p(\bm{x}^i,k^i,i \tau \vert \bm{x}^{i-1},k^{i-1},(i-1)\tau) p_{k^0}(\bm{x}^0,0) \label{path-prob-factor} .
\end{align}
Since computing the KL divergence between two path probabilities involves taking the logarithm of their ratio, it decomposes into a sum of diffusion and jump parts,
\begin{align}
D_\text{KL}(\mathbb{P}^b \Vert \mathbb{P}^a) = D^\text{d}_\text{KL}(\mathbb{P}^b \Vert \mathbb{P}^a) + D^\text{j}_\text{KL}(\mathbb{P}^b \Vert \mathbb{P}^a) + D^0_\text{KL}(p^b_0 \Vert p^a_0).
\end{align}
Here, $\mathbb{P}^b$ and $\mathbb{P}^a$ denote the path probability densities of the two dynamics and the subscripts d, j and $0$ correspond to the contribution from the diffusion- and jump-part and the initial state, respectively.
This allows us to consider the modifications of the diffusive and jump dynamics and their contributions to the relative entropy separately.
We now construct another jump-diffusion process by changing the drift vector and transition rates according to
\begin{subequations}
\begin{align}
\bm{a}_k(\bm{x},t) \ &\rightarrow \ \bm{b}_k(\bm{x},t) = \bm{a}_k(\bm{x},t) + \bm{z}_k(\bm{x},t) \\
W^a_{k l}(\bm{x},t) \ &\rightarrow \ W^b_{k l}(\bm{x},t) = W^a_{k l}(\bm{x},t) e^{Z_{k l}(\bm{x},t)},
\end{align}\label{modification-jd}%
\end{subequations}
which corresponds to adding additional generalized forces and rescaling the transition rates.
It was shown in Ref.~\cite{Dec17} that the transformation of the drift vectors and initial state yields a finite KL divergence between the path probabilities (see also Ref.~\cite{Gir60} for a more rigorous mathematical discussion).
Generalizing this part of the transformation to the jump-diffusion case is straightforward and yields
\begin{subequations}
\begin{align}
D_\text{KL}^\text{d}(\mathbb{P}^b \Vert \mathbb{P}^a) &= \frac{1}{4} \int_0^\mathcal{T} dt \sum_{k} \int d\bm{x} \ \bm{z}_k(\bm{x},t) \big(\bm{B}_k\big)^{-1}(\bm{x},t) \bm{z}_k(\bm{x},t) p^b_k(\bm{x},t) = \frac{1}{4} \int_0^\mathcal{T} dt \ \Av{\bm{z} \bm{B}^{-1} \bm{z}}^{b}_t \\
D_\text{KL}^0(p^b_0 \Vert p^a_0) &= \sum_{k} \int d\bm{x} \ p^b_k(\bm{x},0) \ln \bigg(\frac{p^b_k(\bm{x},0)}{p^a_k(\bm{x},0)} \bigg) = \Av{\ln \frac{p^b_0}{p^a_0}}_0^{b},
\end{align}%
\end{subequations}
for the contributions due to the modifications of the drift vector and initial density, respectively, where we defined the average $\av{\ldots}^{b}$ with respect to the modified dynamics.
Here $p^b_k(\bm{x},t)$ is the solution of the Fokker-Planck-Master equation \eqref{sup-fpe} with the modified drift vector, transition rates and initial density.
We now compute the contribution stemming from the transformation of the transition rates.
Since the path probability factorizes according to \eqref{path-prob-factor}, it is enough to consider the contribution of a short interval $[t,t+\tau]$,
\begin{align}
dD^\text{j}_\text{KL}(\mathbb{P}^b \Vert \mathbb{P}^a) =  \sum_{k, m} \int d\bm{y} \ p^{b,(\bm{y})}(k,t+\tau \vert m,t) p^b_m(\bm{y},t) \ln \Bigg( \frac{p^{b,(\bm{y})}(k,t+\tau \vert m,t)}{p^{a,(\bm{y})}(k,t+\tau \vert m,t)} \Bigg) .
\end{align}
We split the double sum into the terms for $k = n$ and the terms for $k \neq m$, using the explicit expression for the short-time jump propagator \eqref{prop-jd},
\begin{align}
dD^\text{j}_\text{KL}(\mathbb{P}^b \Vert \mathbb{P}^a) = &\sum_m \int d\bm{y} \ \Big(1 - \tau \sum_{l \neq m} W^a_{l m}(\bm{y},t) e^{Z_{l m}(\bm{y},t)} \Big) \ln \Bigg( \frac{1 - \tau \sum_{l \neq m} W^a_{l m}(\bm{y},t) e^{Z_{l m}(\bm{y},t)}}{1 - \tau \sum_{l \neq m} W^a_{l m}(\bm{y},t)} \Bigg) p^b_m(\bm{y},t) \\
&+ \sum_{k,m; k \neq m} \int d\bm{y} \ \tau W_{k m}(\bm{y},t) e^{Z_{k m}(\bm{y},t)} Z_{k m}(\bm{y},t) p^b_m(\bm{y},t) \n .
\end{align}
Expanding to first order in $\tau$ and writing in terms of the modified transition rates $\tilde{W}$, this simplifies to
\begin{align}
dD^\text{j}_\text{KL}(\mathbb{P}^b \Vert \mathbb{P}^a) &= \tau \sum_{k,m; k \neq m} \int d\bm{y} \ \Big( Z_{k m}(\bm{y},t) + e^{-Z_{k m}(\bm{y},t)} - 1 \Big) W^b_{k m}(\bm{y},t) p^b_m(\bm{y},t) + O(\tau^2) .
\end{align}
Thus we find for the jump contribution to the KL divergence
\begin{align}
D^\text{j}_\text{KL}(\mathbb{P}^b \Vert \mathbb{P}^a) = \int_0^\mathcal{T} dt \ \Av{Z + e^{-Z} - 1}^{b}_t .
\end{align}
The total KL divergence between the path densities of the modified and original dynamics is thus given by
\begin{align}
D_\text{KL}(\mathbb{P}^b \Vert \mathbb{P}^a) = \int_0^\mathcal{T} dt \ \bigg(\frac{1}{4} \Av{\bm{z} \bm{B}^{-1} \bm{z}}^{b}_t + \Av{Z + e^{-Z} - 1}_t^{b} \bigg) + \Av{\ln \frac{p^b_0}{p^a_0}}_0^{b} \label{rel-entropy-jd} ,
\end{align}
where the three contributions originate in the modification of the drift vector, transition rates and initial density, respectively.
Each term is positive and vanishes only if the corresponding modification vanishes.
The explicit expression \eqref{rel-entropy-jd} gives meaning to the relative entropy between the path measures in terms of averages of (in principle) measurable quantities.
If each of the three modifications is small, 
\begin{align}
\bm{z}_k(\bm{x},t) = \epsilon \bm{\alpha}_k(\bm{x},t), \qquad Z_{k l}(\bm{x},t) = \epsilon \mathcal{Z}_{k l}(\bm{x},t), \qquad p^b_k(\bm{x},0) = p^a_k(\bm{x},0) + \epsilon q_k(\bm{x}),
\end{align}
then \eqref{rel-entropy-jd} can be expanded in terms of $\epsilon$
\begin{align}
D_\text{KL}(\mathbb{P}^b \Vert \mathbb{P}^a) = \epsilon^2 \Bigg( \int_0^\mathcal{T} dt \ \bigg(\frac{1}{4} \Av{\bm{\alpha} \bm{B}^{-1} \bm{\alpha}}^{b}_t + \frac{1}{2} \Av{\mathcal{Z}^2}_t^{b} \bigg) + \frac{1}{2} \Av{\frac{q^2}{(p^a_0)^2}}^a_0 \Bigg) + O(\epsilon^3) .
\end{align}
If we further assume that the dynamics permits a linear response treatment, i.~e.~$\av{r}^b = \av{r}^a + O(\epsilon)$ for the relevant observables $r$, then we can take the averages with respect to the unmodified dynamics up to leading order,
\begin{align}
D_\text{KL}(\mathbb{P}^b \Vert \mathbb{P}^a) = \epsilon^2 \Bigg( \int_0^\mathcal{T} dt \ \bigg(\frac{1}{4} \Av{\bm{\alpha} \bm{B}^{-1} \bm{\alpha}}^a_t + \frac{1}{2} \Av{\mathcal{Z}^2}^a_t \bigg) + \frac{1}{2} \Av{\frac{q^2}{(p^a_0)^2}}^a_0 \Bigg) + O(\epsilon^3) = \Av{C} + O(\epsilon^3) .
\end{align}
Here $C$ is a positive quantity, which depends only on the modification of the dynamics and is averaged with respect to the unmodified dynamics.

\section{Entropy production and uncertainty relation for Markov jump processes}
In the main text, we showed that the FRI can be used to re-derive the thermodynamic uncertainty relation for Langevin dynamics.
We will now show that the same is true for a Markov jump process.
For simplicity, we focus on a pure jump process in a steady state with occupation probabilities $p_k^{a,\text{st}}$ determined by
\begin{align}
\sum_l \Big(W^a_{k l} p^{a,\text{st}}_l - W^a_{l k} p^{a,\text{st}}_k \Big) = 0.
\end{align}
We assume that the dynamics is irreducible and that the transition rates transform as $W^a_{k l} \rightarrow W^a_{l k}$ under time-reversal.
Then, the steady state is unique and the entropy production rate is given by
\begin{align}
\sigma^\text{st} = \sum_{k, l} W^a_{k l} p_l^{a,\text{st}} \ln \bigg( \frac{W^a_{k l} p_l^{a,\text{st}}}{W^a_{l k} p_k^{a,\text{st}}} \bigg) .
\end{align}
Similar to the Langevin case discussed in the main text, we now want to find a small perturbation of the transition rates $W^b_{k l} = W^a_{k l} e^{\epsilon \mathcal{Z}_{k l}}$ for which the change in a generalized current is proportional to the steady state current $\av{r}^{b} - \av{r}^a = \epsilon \Av{r}^a$ and the relative entropy can be identified with the entropy production.
We make the choice
\begin{align}
\mathcal{Z}_{k l} &= \frac{W^a_{k l} p_l^{a,\text{st}} - W^a_{k l} p_l^{a,\text{st}}}{W^a_{k l} p_l^{a,\text{st}} + W^a_{k l} p_l^{a,\text{st}}},
\end{align}
leading to the modified transition rates
\begin{align}
W^b_{k l} = W^a_{k l} \Bigg( 1 + \epsilon \frac{W^a_{k l} p_l^{a,\text{st}} - W^a_{k l} p_l^{a,\text{st}}}{W^a_{k l} p_l^{a,\text{st}} + W^a_{k l} p_l^{a,\text{st}}} \Bigg) + O(\epsilon^2) .
\end{align}
The corresponding steady state is determined by (up to linear order in $\epsilon$)
\begin{align}
(1+\epsilon) \sum_l \Big(W^a_{k l} p_l^{b,{\text{st}}} - W^a_{l k} p_k^{b,{\text{st}}} \Big) = 0.
\end{align}
It is easy to see that this is solved by $p_k^{b,\text{st}} = p_k^{a,\text{st}}$, i.~e.~the above modification does not change the steady state to leading order.
It is immediately apparent from the above that the jump current $j^\text{j}$ defined in \eqref{sup-fpe} changes as $j^{b,\text{st}} = (1 + \epsilon) j^{a,\text{st}}$.
Defining a generalized current in terms of its average
\begin{align}
\av{r}^a = \int_0^\mathcal{T} dt \ \sum_{k, l} \psi_{k l} j_{k l}(t) = \mathcal{T} \sum_{k, l} \psi_{k l} j_{k l}^\text{st}
\end{align}
we thus have $\av{r}^{b} = (1+\epsilon) \av{r}^a$, just as desired.
Finally, we evaluate the KL divergence
\begin{align}
D_\text{KL}(\mathbb{P}^b \Vert \mathbb{P}^a) = \frac{\mathcal{T} \epsilon^2}{2} \Av{\mathcal{Z}^2}^{a,\text{st}} = \frac{\mathcal{T} \epsilon^2}{2} \sum_{k,l} \Bigg( \frac{W^a_{k l} p_l^{a,\text{st}} - W^a_{k l} p_l^{a,\text{st}}}{W^a_{k l} p_l^{a,\text{st}} + W^a_{k l} p_l^{a,\text{st}}} \Bigg)^2 W^a_{k l} P_l^{a,\text{st}} = \frac{\mathcal{T} \epsilon^2}{4} \sum_{k,l} \frac{\big(W^a_{k l} p_l^{a,\text{st}} - W^a_{k l} p_l^{a,\text{st}} \big)^2}{W^a_{k l} p_l^{a,\text{st}} + W^a_{k l} p_l^{a,\text{st}}} .
\end{align}
We now use the log-mean inequality
\begin{align}
(b-a)(\ln(b)-\ln(a)) \geq \frac{2 (b-a)^2}{b + a} \label{lm-ineq},
\end{align}
which holds for arbitrary $a,b > 0$.
This can be shown using the following argument: for $\beta > \alpha > 0$ we write
\begin{align}
\frac{1}{\beta - \alpha} (\ln(\beta) - \ln(\alpha)) = \frac{1}{\beta-\alpha} \int_\alpha^\beta dx \ \frac{1}{x} &\geq \bigg( \frac{1}{\beta-\alpha} \int_\alpha^\beta dx \ x \bigg)^{-1} = \frac{2 (\beta-\alpha)}{\beta^2 - \alpha^2} \nn
\Rightarrow \quad \ln(\beta) - \ln(\alpha) &\geq 2 \frac{\beta - \alpha}{\beta + \alpha},
\end{align}
where we used the convexity of $1/x$ for positive arguments.
Multiplying both sides by $\beta - \alpha$, we obtain the desire inequality.
Using \eqref{lm-ineq}, we can bound the relative entropy from above
\begin{align}
D_\text{KL}(\mathbb{P}^b \Vert \mathbb{P}^a) &\leq \frac{\mathcal{T} \epsilon^2}{8} \sum_{k,l} \big(W^a_{k l} p_l^{a,\text{st}} - W^a_{l k} p_k^{a,\text{st}}\big) \ln \bigg(\frac{W^a_{k l} p_l^{a,\text{st}}}{W^a_{l k} p_k^{a,\text{st}}} \bigg) = \frac{\mathcal{T} \epsilon^2}{4} \sum_{k,l} W^a_{k l} p_l^{a,\text{st}} \ln \bigg(\frac{W^a_{k l} p_l^{a,\text{st}}}{W^a_{l k} p_k^{a,\text{st}}} \bigg) = \frac{\epsilon^2}{4} \Delta S,
\end{align}
where $\Delta S$ is the entropy production for a steady-state jump process.
We now use the general bound derived in the main text,
\begin{align}
\big(\av{r}^{b} - \av{r}^a \big)^2 \leq 2 \Av{\Delta r^2}^a D_\text{KL}^{b \Vert a} \label{sup-fri},
\end{align}
and obtain the uncertainty relation for the a steady-state Markov jump process,
\begin{align}
2\Av{r}^2 \leq \Av{\Delta r^2} \Delta S .
\end{align}
The corresponding result for a diffusion process with internal states readily follows from the results for Langevin and jump dynamics, since both the KL divergence and the entropy production can be written as a sum of jump and diffusion parts.

\section{Modification of the diffusion matrix}
While the transformation \eqref{modification-jd} accounts for a wide range of physical perturbations, it excludes one important case, which is a change in the diffusion matrix.
The latter is physically relevant because it describes the response of a system to a change in temperature.
Focusing on the diffusion part of the dynamics, we write the Fokker-Planck equation for the modified probability density as
\begin{align}
\partial_t p^b(\bm{x},t) = - \bm{\nabla}\Big(&\bm{a}(\bm{x},t) - \bm{\nabla} \bm{B}(\bm{x},t)  - \bm{\nabla} \bm{D}(\bm{x},t) \Big) p^b(\bm{x},t) ,
\end{align}
where we included a change in the diffusion matrix according to $\bm{B}(\bm{x},t) \rightarrow \tilde{\bm{B}}(\bm{x},t) = \bm{B}(\bm{x},t) +  \bm{D}(\bm{x},t)$.
We remark that this modification of the dynamics violates the condition of absolute continuity between the original and modified process.
As a consequence, the KL divergence between the path densities is infinite.
However, if the change in the diffusion matrix is small, $\bm{D}(\bm{x},t) = \epsilon \bm{\mathcal{D}}(\bm{x},t)$, then by assumption of linear response, we have $p^b(\bm{x},t) = p^a(\bm{x},t) + O(\epsilon)$, and this is to leading order equivalent to
\begin{align}
\partial_t p^b(\bm{x},t) = - \bm{\nabla} \Big(\bm{b}(\bm{x},t) - \bm{\nabla}\bm{B}(\bm{x},t) \Big) p^b(\bm{x},t) \label{fpe-temp-mod},
\end{align}
with the modified drift vector $\bm{b}(\bm{x},t) = \bm{a}(\bm{x},t) - \epsilon [\bm{\nabla}\bm{\mathcal{D}}(\bm{x},t) p^a(\bm{x},t)]/p^a(\bm{x},t)$.
The effect of a change in the diffusion matrix can thus to leading order be represented as a change in the drift vector and we get a bound on the change of the observable due to this modification
\begin{align}
\big( \av{r}^{b} - \av{r}^a \big)^2 \leq 2 \Av{\Delta r^2}^a \av{C}^a \label{fri-diff}
\end{align}
where $\av{C}^a$ is given by
\begin{align}
\av{C}^a = \int_0^\mathcal{T} dt \ \Av{\frac{[\bm{\nabla} \bm{D} p^a] \bm{B}^{-1} [\bm{\nabla} \bm{D} p^a]}{(p^a)^2} }^a_t \label{cost-diffusion} .
\end{align}
We stress that \eqref{fpe-temp-mod} correctly describes only the one-point density $p^b(\bm{x},t)$ to leading order and \emph{not} the transition or path density. 
Thus, the bound \eqref{fri-diff} only holds for observables that depend only on the instantaneous state of the system, i.~e.~whose average can be written as
\begin{align}
\av{r}^a = \int_0^\mathcal{T} dt \int d\bm{x} \ \Big( \phi(\bm{x},t) p^a(\bm{x},t) + \bm{\chi}(\bm{x},t) \bm{j}^a(\bm{x},t) \Big)
\end{align}
with some functions $\phi$ and $\bm{\chi}$.

\section{LFRI for the variance}
The LFRI (Eq.~(12) of the main text) limits the change of the average of an observable $r$ under a small change of the underlying probability distribution.
A natural question is whether a similar bound holds for higher order moments of $r$, in particular for the variance of $r$, i.~e.~how much the variance of an observable can change.
While the variance of $r(\omega)$ is not an observable for the distribution $p^a(\omega)$, i.~e.~there is no $q(\omega)$ independent of the choice of $p^a$ such that $\av{q}^a = \av{\Delta r^2}^a$, it is an observable in the trivially extended probability space
\begin{align}
\av{\Delta r^2}^a &= \int d\omega \int d\omega' \ q(\omega,\omega') p^a(\omega,\omega') = \av{q}^A \\
 \quad \text{with} \quad p^A(\omega,\omega') &= p^a(\omega) p^a(\omega') \quad \text{and} \quad q(\omega,\omega') = r(\omega)^2 - r(\omega) r(\omega') \n .
\end{align}
Since $\omega$ and $\omega'$ correspond to two independent copies of the stochastic process defining $p^a$, the additivity of the KL divergence
\begin{align}
D_\text{KL}^{B \Vert A} = 2 D_\text{KL}^{b \Vert a} .
\end{align}
Then applying the LFRI to $q$, we obtain
\begin{align}
\Big( \Av{\Delta r^2}^b - \Av{\Delta r^2}^a \Big)^2 \leq 4 \Big( \Av{\Delta r^4}^a + 2 \big( \Av{\Delta r^3}^a + \Av{\Delta r^2}^a \Av{r}^a \big) \Av{r}^a \Big) D_\text{KL}^{b \Vert a}
\end{align}
Since the left-hand side is independent of a constant shift $r(\omega) \rightarrow r(\omega) + r_0$, we can minimize the right-hand side with respect to $r_0$ and obtain
\begin{align}
\big( \Av{\Delta r^2}^b - \Av{\Delta r^2}^a \big)^2 \leq 4 \Bigg( \Av{\Delta r^4}^a - \frac{\big(\Av{\Delta r^3}^a\big)^2}{2 \Av{\Delta r^2}^a} \Bigg) D_\text{KL}^{b \Vert a},
\end{align}
which, as expected, is independent of $\av{r}^a$.
This can be written in the compact form
\begin{align}
\Bigg( \frac{\Av{\Delta r^2}^b - \Av{\Delta r^2}^a}{\Av{\Delta r^2}^a} \Bigg)^2 \leq 4 \bigg( \kappa_r^a - \frac{1}{2} (\gamma_r^a)^2 \bigg) D_\text{KL}^{b \Vert a},
\end{align}
where $\gamma_r^a$ and $\kappa_r^a$ are the skewness and the kurtosis, respectively, of the distribution of $r$ in system $a$, $p_r^a(r) = \int_\Omega d\omega \ \delta(r-r(\omega)) p^a(\omega)$.
Thus, while the bound on the change of average is given by the variance in the unperturbed system, the bound on the change of the variance is given by higher-order moments of the unperturbed dynamics.

\vspace{1cm}

\twocolumngrid

% Bibliography
\bibliography{bib}

\end{document}